\documentclass[12pt]{article}
\usepackage{amsmath,amssymb,mathrsfs,url,graphicx}
\usepackage{color}

\title{Bohmian Trajectories For a\\ Time Foliation with Kinks}

\author{
Ward Struyve\footnote{Departments of Mathematics and Philosophy,
     Rutgers University, Hill Center, 
     110 Frelinghuysen Road, Piscataway, NJ 08854-8019, 
     USA. E-mail: wstruyve@math.rutgers.edu}\ \ and 
Roderich Tumulka\footnote{Department of Mathematics,
     Rutgers University, Hill Center, 
     110 Frelinghuysen Road, Piscataway, NJ 08854-8019, 
     USA. E-mail: tumulka@math.rutgers.edu}
}

\date{November 14, 2013}

\newcommand{\be}{\begin{equation}}
\newcommand{\ee}{\end{equation}}
\newcommand{\foliation}{\mathscr{F}} 
\newcommand{\M}{\mathscr{M}} 
\newcommand{\C}{\mathscr{C}} 
\newcommand{\U}{\mathscr{U}} 
\newcommand{\SC}{\mathscr{S}} 

\newcommand{\RRR}{\mathbb{R}}
\newcommand{\CCC}{\mathbb{C}}

\begin{document}
\maketitle
\begin{abstract}
This paper concerns the hypersurface Bohm--Dirac model, i.e., the version of Bohmian mechanics in a relativistic space-time proposed by D\"urr et al.~\cite{HBD}, which assumes a preferred foliation of space-time into spacelike hypersurfaces (called the time foliation) as given. We show that the leaves of the time foliation do not have to be smooth manifolds but can be allowed to have kinks. More precisely, we show that, also for leaves with kinks, the trajectories are still well defined and the appropriate $|\psi|^2$ distribution is still equivariant, so that the theory is still empirically equivalent to standard quantum mechanics. This result applies to the case where the time foliation is determined by the previously proposed law $dn=0$, since such a foliation generically has kinks.

\medskip

\end{abstract}

\section{Introduction}

\begin{figure}[h]
\begin{center}
\includegraphics[width=.7 \textwidth]{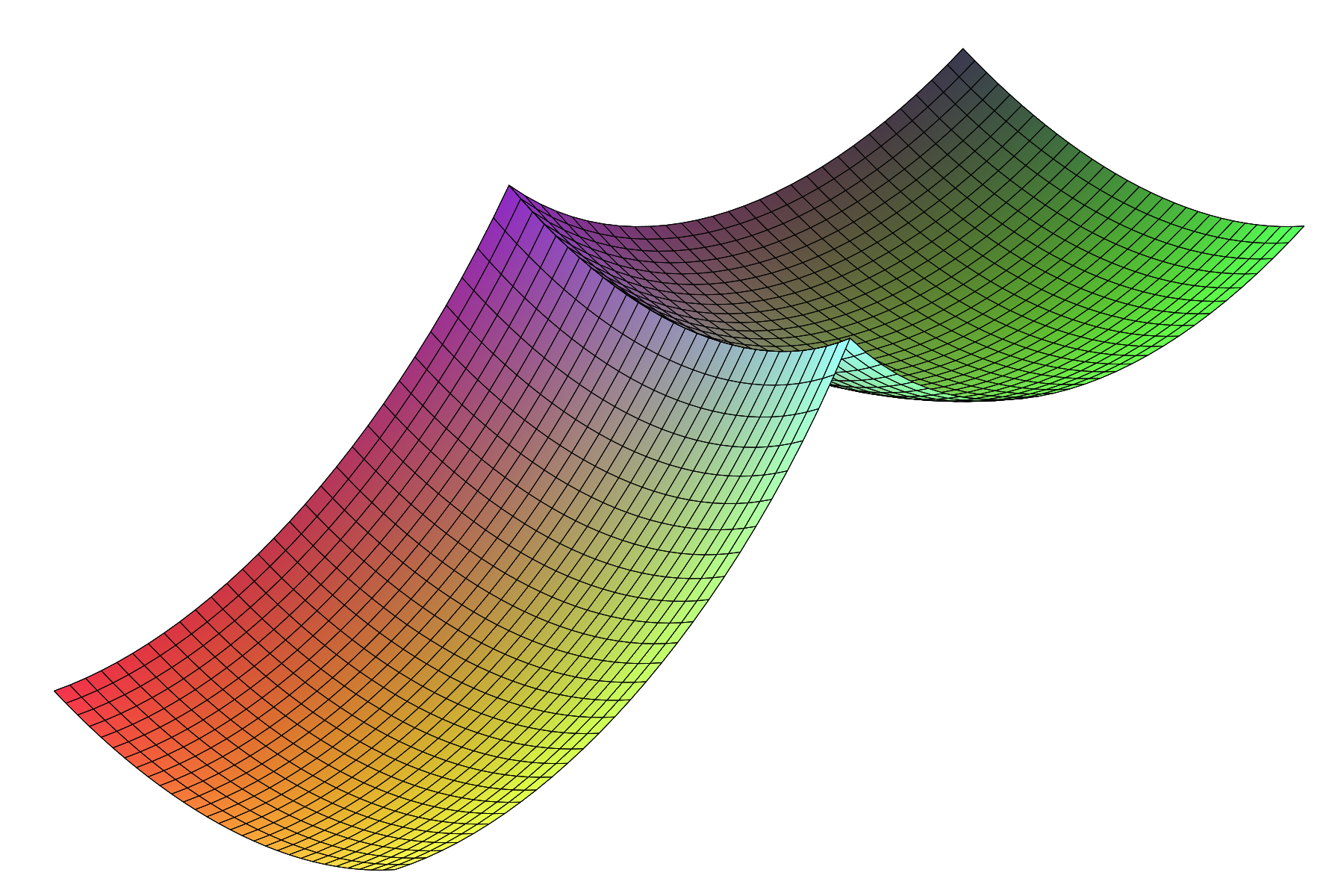}
\end{center}
\caption{An example of a piece of a 2-dimensional surface-with-kinks in 3-dimensional space.}
\label{figone}
\end{figure}

For defining a version of Bohmian mechanics in a relativistic space-time, it seems necessary to assume a preferred slicing of space-time into spacelike hypersurfaces (the ``time foliation'' $\foliation$) \cite{Tum07,GSZ13}. If a time foliation is granted, there is a natural and convincing version of Bohmian mechanics for $N$ particles, known as the hypersurface Bohm--Dirac model (HBDM) \cite{HBD}, which is empirically equivalent to standard quantum mechanics (see Section~\ref{sec:def} for its definition). The set of configurations that are simultaneous with respect to $\foliation$ is denoted by
\be\label{Cdef}
\C= \bigcup_{\Sigma\in\foliation} \Sigma^N\,.
\ee
For the sake of simplicity and definiteness we assume that the particles are non-interacting; introducing interaction does not change the situation in a relevant way, as elucidated in Remark~\ref{rem:noninteract} in Section~\ref{sec:remarks} below. It is usually assumed that the hypersurfaces $\Sigma\in\foliation$ (the ``time leaves'') are smooth. In this paper, we are interested in how the HBDM fares when the time leaves are not smooth but have kinks as in Figure~\ref{figone} (while the wave function is taken to be smooth). In the following, we will use words like ``hypersurface'' and ``foliation'' in the sense that they do not have to be smooth but may involve kinks.

The question we address is whether the HBDM still works for a time foliation with kinks, and this means two things: (a)~whether the trajectories are still well defined, and (b)~whether the appropriate $|\psi|^2$ distribution is still equivariant (i.e., whether the Bohmian evolution from one time leaf to another preserves $|\psi|^2$). 

The motivation for these questions comes from a law governing the time foliation that one of us proposed \cite{Tum07}. According to this law, the foliation is determined by the requirement that all points on a leaf have the same timelike distance (in terms of the space-time metric) to a given ``initial'' hypersurface. (In other words, the ``lapse function'' is constant.) As we discuss elsewhere \cite{ST13a}, the foliation thus defined will generically have kinks, even if the initial hypersurface is smooth. The future-pointing unit normal vector on the foliation is only defined outside the kinks. Denoting its components by $n^\mu$, we have that 
\be\label{dn0}
dn=0\,,
\ee
where $n=n_\mu dx^\mu$ is the one-form with components $n_\mu$, and $d$ is the exterior derivative of a differential form. In the following, we will simply call this law ``$dn=0$'' even though this equation is, strictly speaking, not fully equivalent to the law, since, in particular, the equation does not apply at kinks.

In order to address the questions (a) and (b) above, it is useful to consider the analogous questions in a wider class of dynamical laws including the HBDM, namely the class of random trajectory models (RTMs) defined by a probability current (see Section~\ref{sec:reasoning}). We assume that the time leaves are Cauchy (i.e., intersect every timelike curve exactly once) and have only spacelike (and no lightlike) tangent vectors, including the tangent vectors on both sides of a kink.
With $K$ the kink set (i.e., the set of points in $\C$ where $\Sigma^N$ is not smooth), we will first show that if the condition 
\begin{equation}\label{currentcond1}
\text{the current into }K\text{ from one side} = - \text{the current into }K\text{ from the other side} 
\end{equation}
is satisfied (see Eq.~\eqref{currentcond2} for a detailed formulation), then the trajectories are typically well defined and equivariance will hold. We will then show that this condition
is always satisfied for the HBDM. Thus, the HBDM works just as well in the case of foliations with kinks. In contrast, we show that Slater's law for photon trajectories \cite{Sla75} violates the current condition \eqref{currentcond1} and thus is incompatible with any time foliation with kinks.

The most demanding result is to prove \eqref{currentcond1} for HBDM. The key fact is that the probability current of HBDM can be expressed as a $3N$-form on (space-time)$^N$ that depends on $\psi$ but not on $\foliation$ and thus is continuous on $K$.

We will give a full discussion of foliations defined by $dn=0$ in \cite{ST13a}. As we show there, they have further properties, besides being Cauchy, having only spacelike tangent vectors,  and having kinks, that we do not use here for the proof of equivariance: for example, that the kink set is everywhere timelike (in the sense that its normal, at points where it is defined, is spacelike), that kinks cannot disappear, and that, at any kink point $x\in\Sigma\in\foliation$, the rapidity (i.e., Lorentz-invariant angle) between $\Sigma$ and the kink set in space-time is equal on both sides. Also, isolated points of non-smoothness (such as a conical tip) can form, but generically they do not stay isolated but grow (instantly) into a kink. (And in case they do stay isolated, a Bohmian trajectory has probability zero to ever run into such a point.)

\section{Foliations with Kinks}
\label{sec:def}

Let us introduce some terminology. A \emph{stratified submanifold} is a subset of a smooth manifold
that is the union of several disjoint (pieces of) smooth submanifolds of equal dimension along with their common smooth boundaries. A \emph{submanifold-with-kinks} is a stratified submanifold such that only two manifolds have any codimension-1 boundary in common. That is, $\textsf{Y}$-shaped pieces are allowed for stratified submanifolds but not for submanifolds-with-kinks, and $\wedge$-shaped pieces are allowed for submanifolds-with-kinks but not for smooth submanifolds. As a consequence, submanifolds-with-kinks can be approximated by smooth manifolds but generic stratified submanifolds cannot. A typical example of a submanifold-with-kinks is shown in Figure~\ref{figone}. For any submanifold-with-kinks $S$, the \emph{kink set} $K(S)$ is the set of those $x\in S$ such that $S$ is not a smooth submanifold in any neighborhood of $x$. The kink set is a stratified submanifold with $\dim K(S) = (\dim S)-1$. In particular, the kink set can have kinks as well, and three or more kinks can meet in a corner; in fact, such corners usually occur for hypersurfaces arising from $dn=0$.

\begin{figure}[h]
\begin{center}
\includegraphics[width=.6 \textwidth]{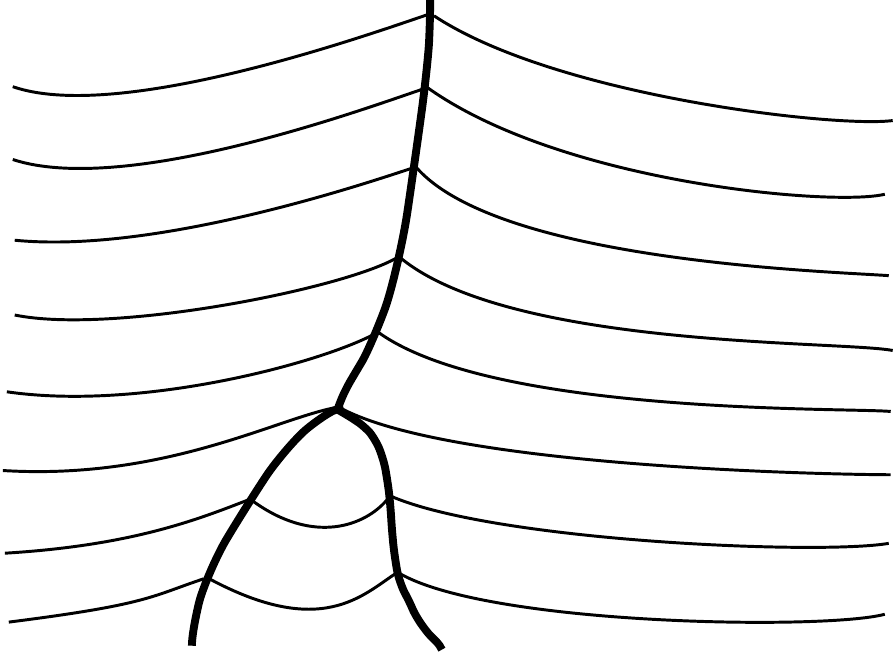}
\end{center}
\caption{An example of a foliation $\foliation$ of $1+1$-dimensional space-time into hypersurfaces with kinks (thin lines). Thick lines: the kink set $K(\foliation)$, which is a stratified submanifold.}
\label{figtwo}
\end{figure}

Let $\M$ denote Minkowski space-time.\footnote{The considerations of this paper work as well in a curved space-time that is globally hyperbolic (i.e., without space-time singularities or closed timelike curves).}  
A foliation-with-kinks $\foliation$ is a foliation whose leaves are submanifolds-with-kinks. We take for granted about $\foliation$ that $K(\foliation)=\cup_{\Sigma\in\foliation} K(\Sigma)$ is itself a stratified submanifold of dimension 3; see Figure~\ref{figtwo} for an example. This is the case for the law $dn=0$.

If $\foliation$ is smooth then $\C$ is a smooth submanifold of $\M^N$; if $\foliation$ has kinks then $\C$ is a submanifold-with-kinks of $\M^N$ (compare to Figure~\ref{figone}). In both cases, $\C$ has dimension $3N+1$. We define
\be\label{KCdef}
K(\C,\foliation) = \bigcup_{\Sigma\in\foliation} K(\Sigma^N)\,,
\ee
where $K(\Sigma^N)$ is the kink set of $\Sigma^N$, i.e.,
\be\label{KSigma}
K(\Sigma^N) = \bigcup_{j=1}^N \Sigma^{j-1}\times K(\Sigma) \times \Sigma^{N-j}\,.
\ee
If $N>1$, we have that $K(\C,\foliation) = K(\C)$, i.e., the kink set of $\C$, which is a $3N$-dimensional stratified submanifold of $\M^N$. If $N=1$, $\C=\M$ and $K(\C)$ is empty, so that $K(\C,\foliation)=K(\foliation) \neq K(\C)$ in the case of a foliation with kinks.

\section{Definition of the Hypersurface Bohm--Dirac Model}

The HBDM for $N$ non-interacting particles \cite{HBD} employs a multi-time wave function $\psi:\M^N\to (\CCC^4)^{\otimes N}$ satisfying a Dirac equation for each particle,\footnote{If space-time is curved, then the wave function $\psi$ is a cross section of the appropriate spin bundle over $\M^N$, and $\partial_{j\mu}$ in \eqref{Dirac} needs to be replaced by a covariant derivative. The equivariance of the HBDM in curved space-time was established in \cite{Tum01}.}
\be\label{Dirac}
\gamma_j^\mu \Bigl(i\hbar\partial_{j\mu}+e_jA_{j\mu}(x_j)\Bigr) \psi = m_j \psi\,,  
\ee
where $e_j$ and $m_j$ are charge and mass of the $j$-th particle, $\psi=\psi(x_1,\ldots,x_N)$, $\partial_{j\mu} = \partial/\partial x_j^\mu$, and
\be
\gamma_j^\mu = I^{\otimes (j-1)}\otimes \gamma^\mu \otimes I^{\otimes (N-j)}
\ee
with $I$ the identity matrix and $\gamma^\mu$ the Dirac matrices.

Consider a smooth time foliation $\foliation$ and let $n^\mu(x)$ denote its future unit normal vector field.
For $j=1,\ldots,N$, let $X^\mu_j(s)$ be any parameterization of the world line of particle $j$. The world lines of the HBDM are everywhere timelike-or-lightlike. Since every time leaf $\Sigma\in\foliation$ is a Cauchy hypersurface, every world line $X_j(\cdot)=X_j^\mu(\cdot)$ intersects it exactly once, at $s(\Sigma)$; we write $X_j(\Sigma)$ for $X_j(s(\Sigma))$, and $X(\Sigma)$ for the configuration $\bigl(X_1(\Sigma),\ldots, X_N(\Sigma)\bigr)$. The equation of motion reads
\be\label{Bohm}
\frac{dX_j^\mu}{ds}(s(\Sigma))\propto \Bigl(\overline{\psi} \bigl[\gamma^{\mu_1}\otimes \cdots \otimes \gamma^{\mu_N}\bigr] \psi\Bigr)(X(\Sigma)) \, \delta^\mu_{\:\:\mu_j} \, \prod_{k\neq j} n_{\mu_k}(X_k(\Sigma))\,.
\ee
The right-hand side is always timelike-or-lightlike.

Any Cauchy hypersurface $\Sigma$ (not necessarily from $\foliation$) defines a $3N$-dimensional configuration space $\Sigma^N$. By the ``$|\psi|^2$ distribution on $\Sigma^N$'' we mean the probability distribution on $\Sigma^N$ that has density, relative to the measure $d^3x_1 \cdots d^3x_N$ with $d^3x$ the invariant (Riemannian) volume measure on $\Sigma$,
\be
\rho_{\Sigma}(x_1,\ldots,x_N) = 
\Bigl(\overline{\psi} \bigl[\gamma^{\mu_1}\otimes \cdots \otimes \gamma^{\mu_N}\bigr] \psi\Bigr)
(x_1,\ldots,x_N) \, \prod_{k=1}^N n_{\mu_k}(x_k)
\ee
for any $x_1,\ldots,x_N\in\Sigma$, where $n^\mu(x)$ denotes the future unit normal vector to $\Sigma$ at $x\in\Sigma$.

The equivariance theorem asserts that if on some hypersurface $\Sigma\in\foliation$ the configuration $X(\Sigma)$ is random with $|\psi_{\Sigma}|^2$ distribution, then the same is true for any other $\Sigma\in\foliation$ (but not necessarily for $\Sigma \notin \foliation$). This follows (but not immediately) from
\be\label{divergence}
\partial_{j\mu_j}\Bigl(\overline{\psi} \bigl[\gamma^{\mu_1}\otimes \cdots \otimes \gamma^{\mu_N}\bigr] \psi\Bigr)=0\,,
\ee
which in turn is a consequence of the Dirac equation \eqref{Dirac} \cite{HBD,Tum01}.

\section{Main Reasoning}
\label{sec:reasoning}

It will be useful to think in terms of global coordinates on $\C$; by this we mean a homeomorphism $\varphi:\C\to\RRR^{3N+1}$ that is a local diffeomorphism outside of $K(\C,\foliation)$. Furthermore, it will be convenient to choose $\varphi$ in such a way that its time coordinate $\varphi^0$ is constant on the hypersurfaces $\Sigma^N$ for $\Sigma\in\foliation$. In fact, it will be convenient to obtain $\varphi$ from a curvilinear coordinate system $\tilde\varphi:\M\to\RRR^4$ whose time coordinate is constant on time leaves. Also $\tilde{\varphi}$ is a homeomorphism and a local diffeomorphism outside the kink set $K(\foliation)$. The coordinate system $\varphi$ allows us to translate our questions to $\RRR^{3N+1}$.

\subsection{Probability Current in Coordinate Space}

First, ignore the kinks in $\foliation$. One easily sees that, for a given $\psi$, the equation of motion \eqref{Bohm} defines a field of directions on $\C$ (and thus, by $\varphi$, on $\RRR^{3N+1}$), and that any $N$-tuple of world lines satisfying \eqref{Bohm} corresponds to an integral curve of that direction field. Define the \emph{probability current vector field} $j=(j^0,\vec{\jmath})$ on $\RRR^{3N+1}$ to be the vector field that has just these directions, with
\be\label{j0coo}
j^0(\varphi(x_1,\ldots,x_N)) = \prod_{j=1}^N \sqrt{-\det(^{3}g(x_j))} \: \rho_\Sigma(x_1,\ldots,x_N)
\ee
for all $x_1,\ldots,x_N\in\Sigma$, where $^3g$ is the 3-metric on $\Sigma$ in the $\tilde\varphi$ coordinates. The quantity \eqref{j0coo} is the density of the $|\psi_\Sigma|^2$ distribution in the $\varphi$ coordinates (relative to coordinate volume). The properties of $j$ are familiar from non-relativistic Bohmian mechanics: The trajectories $(t,X(t))$ in $\RRR^{3N+1}$ satisfy
\be\label{dXdt}
\frac{dX}{dt} = \frac{\vec{\jmath}}{j^0}(t,X(t))
\ee
and, by \eqref{divergence}, the continuity equation
\be\label{continuity}
\frac{\partial j^0}{\partial t} = -\sum_{k=1}^{3N} \frac{\partial j^k}{\partial x^k}
\ee
holds, implying equivariance of $j^0$, the probability density of $X(t)$.

Now consider the kinks in $\foliation$. The kinks have the effect that $j$ is discontinuous at the (coordinate image of the) kink set $K=\varphi(K(\C,\foliation))$, the union of $3N$-dimensional hypersurfaces in $\RRR^{3N+1}$ with common boundaries. At any other point $\varphi(x_1,\ldots,x_N)$ of $\RRR^{3N+1}$, $j$ is smooth because $\psi$ and $\varphi$ are smooth at $(x_1,\ldots,x_n)$, and $\Sigma\in\foliation$ (and thus $n_\mu$) is smooth at each of $x_1,\ldots,x_N$. The discontinuity is, in fact, a finite jump; more precisely, it is such that $j$ has a limit on $K$ from each side of $K$. That is because the normal vector field $n^\mu$ has a limit at a kink from each side, while $\psi$ and $\varphi$ are continuous.\footnote{We are using our assumption here that every $\Sigma\in\foliation$ has only spacelike tangent vectors (and no lightlike ones); if it had lightlike tangent vectors, say at $x\in\Sigma$, then $n_\mu(x)$ would not be defined; if, in addition, $x$ was a kink of $\Sigma$, then one of the limits of $n_\mu$ on the kink set would not be defined. This assumption is very mild, as it is generic for Cauchy hypersurfaces not to have any lightlike tangent vectors (i.e., by a suitable small perturbation, lightlike tangent vectors can be removed).
Moreover, in the case of foliations $\foliation$ determined by $dn= 0$, we believe it is the case that the tangent vectors of $\Sigma\in\foliation$ are all spacelike, provided that the initial hypersurface is spacelike and Cauchy.}
It will be convenient to call one side of $K$ the left side and the other the right side, even though there is, of course, no consistent rule for selecting one of the sides as the left side, and even though naming the sides will only locally be possible, as $K$ need not be orientable. For $(t,q)\in K\subset \RRR^{3N+1}$, let $j_L(t,q)$ denote the limit of $j$ at $(t,q)$ coming from the left, and $j_R(t,q)$ the limit from the right.

So, the situation of a time foliation with kinks gets translated into a situation with a jump discontinuity of the current vector field $j$ along the hypersurface $K$. More generally, we want to define a \emph{random trajectory model} (RTM) for a given vector field $j$ on $\RRR^{3N+1}$ with the properties 
\begin{enumerate}
\item that $j^0\geq 0$;
\item that $j$ is smooth outside $K$;
\item\label{item:limit} that $j$ on the left side of $K$ possesses locally a smooth continuation to a neighborhood of $K$ (which disagrees, of course, with $j$ on the right side of $K$);\footnote{Given property 2, property 3 is equivalent to $j$ and all of its derivatives having continuous left limits on $K$ \cite{seeley}.} likewise for $j$ on the right side of $K$;
\item that the continuity equation \eqref{continuity} holds outside $K$; 
\item\label{item:vorletzte} and that
\be\label{normalization}
\int_{\RRR^{3N}} d^{3N}x \: j^0(0,x) = 1
\ee
as the initial normalization.
\end{enumerate}
The RTM is then defined to have as possible trajectories the solutions of \eqref{dXdt} (or, equivalently, the integral curves of $j$), and such a probability distribution that the probability density of $X(0)$ is $j^0(0,\cdot)$.\footnote{We ignore here the possibility that trajectories may run to spatial infinity in finite time (which cannot happen for HBDM because there the particles cannot move faster than light), or run into a node of $j$.}

\subsection{Continuation of Trajectories and Current Condition}

\begin{figure}[h]
\begin{center}
\includegraphics[width=.4 \textwidth]{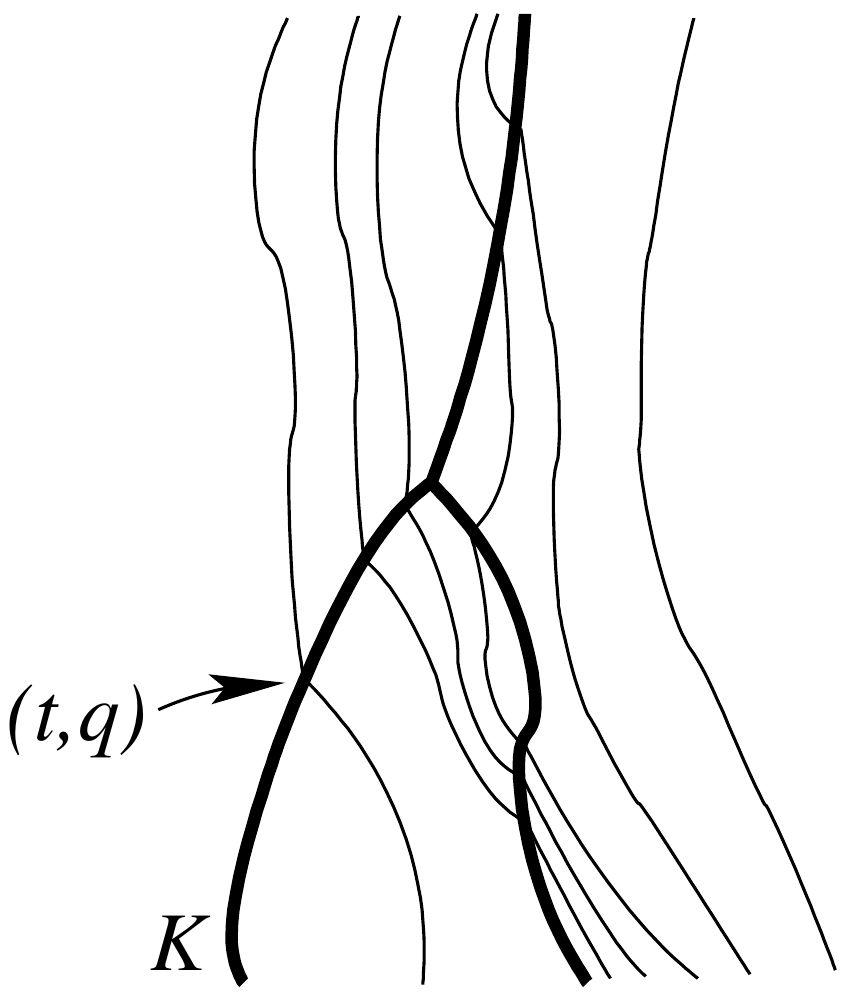}
\end{center}
\caption{The coordinate space $\RRR^{3N+1}$ with the kink set $K$ (thick lines) and several Bohmian trajectories (thin lines) that are integral curves of the current vector field $j$. At $(t,q)\in K$, an incoming trajectory gets connected to an outgoing trajectory. The two-dimensional picture shown corresponds to $N=1$ and $1+1$-dimensional space-time.}
\label{fig:3}
\end{figure}

The further prescription, needed to define the trajectories globally, specifies what happens when a trajectory in $\RRR^{3N+1}$ hits $K$, and the prescription we want is the obvious one: Suppose that, at the point $(t,q)\in K$, exactly one trajectory ends coming from the left (right) and exactly one trajectory begins toward the right (left), then the two trajectories should be connected, so the beginning trajectory is regarded as the continuation of the ending one; see Figure~\ref{fig:3}. In this way, the total trajectory is continuous, while the velocity must be expected to have a jump discontinuity when the trajectory crosses $K$.

Several questions arise about the possibility of this prescription: Is it true that, at every $(t,q)\in K$, exactly one trajectory begins and one ends? And that one lies on the left and the other on the right side of $K$? From Property \ref{item:limit} of $j$ it follows by standard theorems on the existence and uniqueness of solutions of ODEs that every $(t,q)\in K$ with $j_R(t,q)\neq 0 \neq j_L(t,q)$ lies on one trajectory on the left side of $K$ and one on the right. Each of these trajectories either begins or ends at $(t,q)$. If one ends and one begins, the above prescription works. Can it happen that both end, or both begin? Let us postpone this question for a moment and turn to the question of equivariance: Is, for any piece $P\subset K$, the flux coming out of $P$ on one side equal to the flux going in on the other side, as it should? That is, is the ``amount'' of $j^0$ (of probability, if you wish) coming out equal to the amount going in? Or is some amount lost or gained at $P$, as it should not? 

The condition for conservation of probability at $K$ is that at almost every point $(t,q)\in K$, the density of incoming flux equals the density of outgoing flux; equivalently, that the component of the current across $K$ on one side equals the one on the other side. 
That is, the component across $K$ of $j_L(t,q)$ equals that of $j_R(t,q)$. To compute this component, introduce an arbitrary (Euclidean or Lorentzian) scalar product on $\RRR^{3N+1}$; let $n_K(t,q)$ be a (left-pointing or right-pointing) vector that is normal (with respect to the scalar product) on $K$ at $(t,q)$; then $n_K(t,q)\cdot j(t,q)$ (using the same scalar product) is, up to a factor, the component across $K$ of the current vector field $j$. Thus, the condition can be expressed as 
\be\label{currentcond2}
n_K(t,q)\cdot j_L(t,q) = n_K(t,q) \cdot j_R(t,q)\quad \text{at almost all }(t,q)\in K\,.
\ee
``Almost all'' means that exceptions are allowed as long as they form a set of measure zero in the $3N$ dimensions of $K$. Note that changing the sign of $n_K(t,q)$ does not change the content of the condition, and that changing $n_K(t,q)$ by a factor does not either.\footnote{\label{fn:J1}Readers may be surprised that the content of condition \eqref{currentcond2} does not depend on the choice of scalar product. The deeper reason for that is that a current vector field $j$ can be translated into a $3N$-form $\mathscr{J}$ by contracting $j$ with the $3N+1$-dimensional Levi-Civita symbol $\varepsilon^{(3N+1)}_{A_1\ldots A_{3N+1}}$ (with indices $A_i=0,\ldots,3N$); this procedure is similar to, though not exactly the same as, the Hodge star operator (the latter applies to forms, not to vector fields). The Levi-Civita symbol can be regarded as a differential form of degree $3N+1$ (volume form); any scalar product and orientation on a vector space define a volume form, which differs from $\varepsilon^{(3N+1)}$ only by a scalar factor; this factor, as we just said, cancels out of \eqref{currentcond2}. A key property of $\mathscr{J}$ is that the flux across any piece $P$ of oriented $3N$-surface in $\RRR^{3N+1}$ is given by $\int_P \mathscr{J}$, a quantity that is independent of any choice of scalar product; that is, for any scalar product whose volume form is $\lambda \varepsilon^{(3N+1)}$, where $\lambda\in\RRR\setminus\{0\}$, we have that $\int_P \mathscr{J}=\lambda^{-1}\int_P dA\, n_P\cdot j$ with $dA$ the $3N$-dimensional area defined by the scalar product and $n_P$ the unit normal on $P$ (properly oriented). Thus, the quantity $\lambda^{-1} dA\, n_P\cdot j$ is actually independent of the choice of scalar product.} Equation \eqref{currentcond2} is the precise version of \eqref{currentcond1}; the reason why a minus sign appears in \eqref{currentcond1} but not in \eqref{currentcond2} is that if $n_K(t,q)$ points to the right side of $K$ then $n_K(t,q)\cdot j_L(t,q)$ is the current (density at $(t,q)$) into $K$ from the left side and $n_K(t,q)\cdot j_R(t,q)$ is \emph{minus} the current (density at $(t,q)$) into $K$ from the right side (and both signs change if $n_K(t,q)$ points to the left).

Let us return to the question whether the continuation of a trajectory across $K$ is possible, that is, whether of the two trajectories reaching $(t,q)\in K$, one begins and one ends. If the equation \eqref{currentcond2} is satisfied at $(t,q)$, and if both sides of the equation are nonzero, then one trajectory begins and one ends at $(t,q)$. Actually, for this it is sufficient that both sides of \eqref{currentcond2} have the same sign. Specifically, if $n_K(t,q)$ points to the left side and if both sides of the equation are positive, then one trajectory lying to the left of $K$ begins at $(t,q)$ and one trajectory lying to the right of $K$ ends at $(t,q)$; see Figure~\ref{fig:3}; if both sides of the equation are negative, then left and right are interchanged.

If both sides of the equation \eqref{currentcond2} are zero at $(t,q)$ then nothing can be concluded about whether one trajectory begins and one ends there. However, the trajectories ending or beginning at such a $(t,q)$ form a set of probability 0.

To sum up so far, if \eqref{currentcond2} is satisfied, then almost every trajectory in $\RRR^{3N+1}$ can be unambiguously extended across $K$, and probability is conserved in the sense that the distribution of the trajectories at coordinate time $t$ agrees with $j^0(t,\cdot)$---that is, for the HBDM, with the $|\psi|^2$ distribution on $\Sigma^N$.

Before we turn to the proof that \eqref{currentcond2} is always true in the HBDM, we note that Properties 1--\ref{item:vorletzte} above are satisfied for the current field $j$ defined by the HBDM around \eqref{j0coo}: 1 by \eqref{j0coo}; 2 whenever $\psi$ is smooth (because $\Sigma\in \foliation$ is smooth apart from its kinks); 3 whenever $\psi$ is smooth (because the normal field $n_\mu$ on one side of $K(\foliation)$ can be continued smoothly on the other side of $K(\foliation)$ provided that, as assumed, the time leaves never become lightlike); 4 by \eqref{continuity}; and 5 by assumption.

\subsection{Proof of the Current Condition for the HBDM}

To prove \eqref{currentcond2}, we use differential forms. In $4N$-dimensional $\M^N$ ($N>1$), let $\varepsilon$ be the volume form defined by the metric on $\M$; that is, $\varepsilon=\tilde{\varepsilon} \wedge \cdots \wedge \tilde{\varepsilon}$ with $N$ factors, where $\tilde{\varepsilon}$ is the volume form on $\M$ defined by the metric (i.e., $\tilde{\varepsilon}_{0123}=1$ in every properly oriented Lorentz frame). Define a $3N$-form $J$ on $\M^N$ by
\be\label{Jdef}
J_{\Delta_1\ldots \Delta_{3N}} =
\Bigl(\overline{\psi} \bigl[\gamma^{\mu_1}\otimes \cdots \otimes \gamma^{\mu_N}\bigr] \psi \Bigr)\:
\varepsilon_{1\mu_1,\Delta_1,\Delta_2,\Delta_3,2\mu_2,\ldots,N\mu_N,\Delta_{3N-2},\Delta_{3N-1},\Delta_{3N}}
\ee
with $\Delta_i=1,\ldots,4N$, or, equivalently,
\be\label{Jdef2}
J_{\Delta_1\ldots \Delta_{3N}} = (-1)^{N(N-1)/2}
\Bigl(\overline{\psi} \bigl[\gamma^{\mu_1}\otimes \cdots \otimes \gamma^{\mu_N}\bigr] \psi \Bigr)\:
\varepsilon_{1\mu_1,2\mu_2,\ldots,N\mu_N,\Delta_{1},\ldots,\Delta_{3N}}\,.
\ee
Note that $J$ does not depend on $\foliation$ and is smooth if $\psi$ is.

Now, as shown in \cite[Sec.~4.6]{Tum01}, in the HBDM with a smooth time foliation $\foliation$, the probability flux across any piece $P$ of oriented $3N$-surface in $\C$ (i.e., the expected number of signed crossings of the random trajectory through $P$) is given by $\int_P J$.\footnote{The $3N$-form $J$ just defined is more or less the same as the $3N$-form $\mathscr{J}$ considered in Footnote~\ref{fn:J1}. The differences are (i)~that $\mathscr{J}$ was defined in coordinate space, not on $\C\subset \M^N$, and (ii)~that $\mathscr{J}$ was defined only in $3N+1$ dimensions, not in $4N$ dimensions. In fact, $\mathscr{J} = \varphi_*(J|_{\C})$.} For a time foliation $\foliation$ with kinks this fact implies that every point in $\C\setminus K(\C)$ has a neighborhood $\U$ in which the current is given by $J$, i.e., in which the flux across any piece $P$ of oriented $3N$-surface is given by $\int_P J$. That is, the \emph{current form} is given by $J$ at any point in $\C\setminus K(\C)$. The current form $J$ on $\C\setminus K(\C)$ is related to the current vector field $j$ on $\RRR^{3N+1}\setminus K$ according to
\be\label{Jj}
(\varphi_*J)_{A_1\ldots A_{3N}} = j^{A_0} \: \varepsilon^{(3N+1)}_{A_0, A_1\ldots A_{3N}} \,,
\ee
where $\varphi_* J$ denotes the coordinate expression for $J$ in the coordinate system $\varphi$, $A_i=0,\ldots,3N$, and $\varepsilon^{(3N+1)}$ is the Levi-Civita symbol in $3N+1$ dimensions (i.e., the volume form in coordinate space). 

\begin{figure}[h]
\begin{center}
\includegraphics[scale=.8]{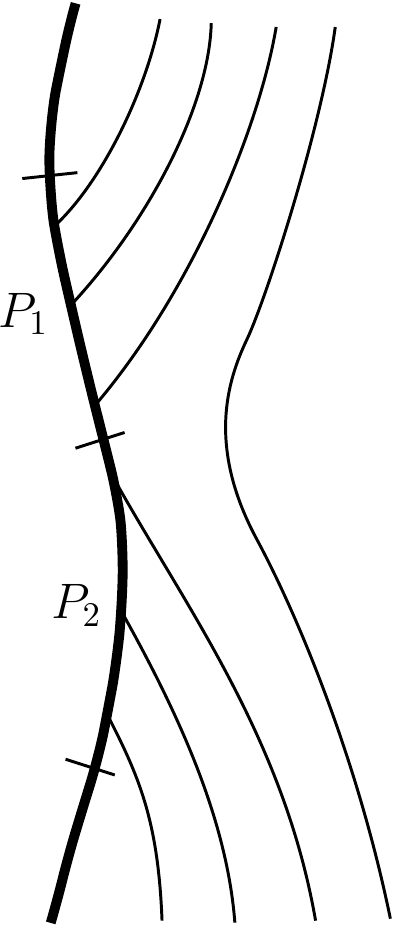}
\end{center}
\caption{Coordinate representation of the submanifold-with-kinks $\C$ with the kink set $K(\C,\foliation)$ (thick line) and several Bohmian trajectories (thin lines) on the right side of $K(\C,\foliation)$. $P \subset K(\C,\foliation)$ is the union of $P_1$ and $P_2$, with $P_1$ the subset of $P$ where trajectories {\em leave} $P$ to the right and $P_2$ the set where trajectories {\em reach} $P$ from the right. The two-dimensional picture shown corresponds to $N=1$ and $1+1$-dimensional space-time.
}
\label{fig:4}
\end{figure}

Now consider, for a piece $P$ of $3N$-surface belonging to $K(\C)$, the flux across $P$ from the right, i.e., the signed ``number'' of trajectories crossing $P$ from the right, which is the ``number'' of trajectories on the right of $P$ ending in $P$ minus the ``number'' starting in $P$, see Figure \ref{fig:4}.
We show that the flux across $P$ from the right is given by $\int_P J$. Since the same argument applies to the flux from the left, it will then follow that the flux density on the right equals almost everywhere the flux density on the left, so that \eqref{currentcond2} is satisfied. The basic reason is that it does not play a role that $P$ lies on the boundary of the ``right half'' of $\C$ because the right half of $\C$ can be extended smoothly.

In more detail, consider a point $\xi=(x_1\ldots x_N)\in K(\C)$ and a ``half-neighborhood'' $\U$ of it in $\C$ on the right side of $K(\C)$ (that is a piece of $3N+1$-dimensional manifold-with-boundary, with $\xi$ lying on the boundary). According to \eqref{KCdef} and \eqref{KSigma}, $\U$ is the union of $\U_\Sigma=\U\cap\Sigma^N$, and, by making $\U$ smaller, $\U_\Sigma$ can be chosen to be a Cartesian product
\be
\U_\Sigma = \tilde\U_{\Sigma,1} \times \cdots \times \tilde\U_{\Sigma,N} 
\ee 
with one $\tilde\U_{\Sigma,j}$ a piece of 3-dimensional manifold-with-boundary in $\M$ (and the boundary lying in $K(\Sigma)$), and all other $\tilde\U_{\Sigma,k}$ ($k\neq j$) being smooth pieces of $\Sigma$ (that are open in $\Sigma$). $\foliation$ could be modified into $\foliation'$ in such a way that $\foliation'$ is smooth and each $\tilde\U_{\Sigma,k}$ ($k=1,\ldots,N$) is contained in some $\Sigma'\in\foliation'$; in other words, the piece $\tilde\U_{\Sigma,j}$ can always be extended smoothly beyond the kink. The signed number of trajectories in $\C$ crossing $P\subset \U\cap K(\C)$ from the right equals the signed number of trajectories in $\C'$ crossing through $P$, and the latter is given by $\int_P J$. 
This completes the proof.\footnote{Eq.~\eqref{Jj} may easily suggest the following incorrect argument. By continuity, \eqref{Jj} implies that for any $(t,q)\in K$,
$j^{A_0}_L(t,q)  \: \varepsilon^{(3N+1)}_{A_0, A_1\ldots A_{3N}} =
\bigl(\varphi_*J(\varphi^{-1}(t,q))\bigr)_{A_1\ldots A_{3N}} =
j^{A_0}_R(t,q)  \: \varepsilon^{(3N+1)}_{A_0, A_1\ldots A_{3N}}$
and thus $j_L(t,q) = j_R(t,q)$. If that were correct, then the trajectory would not even have a kink when crossing $K$. The mistake lies in the fact that $\varphi_*J(\varphi^{-1}(t,q))$ is actually not defined, although $J(\varphi^{-1}(t,q))$ is; the latter is an element of $\Lambda^{3N}(T_{x_1}\M \times \cdots \times T_{x_N}\M)$ with $(x_1\ldots x_N)=\varphi^{-1}(t,q)$, but in order to restrict a $3N$-form on $4N$-space to a $3N+1$-dimensional subset one needs to know the tangent space of the subset, and $\varphi^{-1}(t,q)$ lies on the kink set $K(\C)$ where the tangent space is not uniquely defined; rather, there is a subspace tangent on the right side and a different one that is tangent on the left.}

\section{Remarks}
\label{sec:remarks}

\begin{enumerate}
\item The extended trajectory (i.e., the concatenation of the trajectory ending at $(t,q)\in K$ and the trajectory beginning there) is continuous but usually not differentiable at the crossing point $(t,q)\in K$: it has a kink there. For the world lines of the $N$ particles in space-time, this means that whenever one particle crosses $K(\foliation)$ (i.e., the kink set in space-time $\M$), say on $\Sigma$, the world lines of all other particles usually have kinks (i.e., jumps of velocity) on $\Sigma$. Curiously, the particle crossing $K(\foliation)$ does not have a kink in its world line because, as can be seen from \eqref{Bohm}, its velocity depends on the $n_\mu$ vectors at the locations $X_k(\Sigma)$ of the other particles but not on $n_\mu$ at its own location (and the latter is the only one that jumps). This is connected to the fact that the motion of a single particle does not depend on $\foliation$. As a further consequence of this fact (and the form of \eqref{Bohm}), a particle not entangled with other particles does not have kinks in its world line.

\item It is known \cite{DGZ04} that, in a universe governed by Bohmian mechanics, world lines cannot be observed with high accuracy and without changing them. If they could, these kinks would provide a means of empirically determining the time foliation $\foliation$. It is known that $\foliation$ cannot be determined in the HBDM. Moreover, in a Bohmian universe it is impossible to detect any kink in a world line empirically.

\item Kinks in Bohmian trajectories are also known to occur at particle creation and annihilation \cite{crlet}; that is, whenever a particle gets created or annihilated, the configuration jumps in the configuration space of a variable number of particles, and all other entangled particles usually undergo a discontinuous change in their velocity.

\item Slater \cite{Sla75} suggested in 1924 that a photon wave function is mathematically equivalent to a classical Maxwell field $F_{\mu\nu}$ and guides a photon particle according to the Bohm-like equation of motion
\be\label{Slater1}
\frac{dX^\mu}{ds} \propto T^{\mu 0}(X(s))\,,
\ee
where the tensor field $T^{\mu\nu}$ corresponds to the stress--energy--momentum tensor in the classical Maxwell theory.
Given a foliation with future unit normal vector field $n^\mu(x)$, Eq.~\eqref{Slater1} can be generalized to 
\be\label{Slater2}
\frac{dX^\mu}{ds} \propto T^{\mu\nu}(X(s))\, n_{\nu}(X(s))\,.
\ee
(Just as in Slater's original formulation, the 4-velocity is always timelike or lightlike.) Furthermore, the equation of motion can easily be generalized for $N$ particles in a way similar to \eqref{Bohm}.

A problem with Slater's original theory is that the equivariant density $T^{00}$ does not correspond to the photon number density found in experiments with many photons in the same quantum state; rather, it corresponds to energy density, which differs from the photon number density by a factor of $\hbar \omega$ whenever $\omega$ is sharply defined. Another problem, which arises in the general case of the law \eqref{Slater2} (unless $n^\mu$ is a Killing vector field), is that the density $T^{\mu\nu}n_{\mu}n_{\nu}$ will not be equivariant from one time leaf to another, since $j^\mu=T^{\mu\nu}n_\nu$ has nonzero 4-divergence (unless $n^\mu$ is a Killing vector field).

The generalized law \eqref{Slater2} is also incompatible with kinks in $\foliation$. This is connected to the fact that the motion depends on $\foliation$ already in the case of a single particle, as is evident from \eqref{Slater2}. Relatedly, for $N=1$ the $3N$-form $J$ on $\M^N$ depends on $\foliation$ (unlike for the HBDM), as $J_{\lambda\mu\nu}=T^{\rho\sigma}\, n_{\sigma} \, \tilde{\varepsilon}_{\rho\lambda\mu\nu}$. This implies that our proof for the current condition \eqref{currentcond2} that was used for the HBDM does not carry over to this case. As a consequence, we do not have equivariance of the corresponding distribution. But of course we did not even have equivariance in the case the foliation did not have kinks (unless $n^\mu$ is a Killing vector field). What about the trajectories? Can they be continued across the kink surface? Recall that the current condition is sufficient for this, but not necessary. In order to be able to continue trajectories across the kink surface it is sufficient that $n_K \cdot j_L$ and $n_K \cdot j_R$ have the same sign (they need not be equal). It is easy to see that this is not necessarily true in the case of the generalized law \eqref{Slater2}. Namely, if $j^\mu_L \neq j^\mu_R$ at a space-time point $x$, then one can always find a surface such that the normal $n_K$ to that surface at $x$ is such that $n_K \cdot j_L$ and $n_K \cdot j_R$ have opposite signs. If this surface happens to be the kink surface then the trajectories cannot be continued. In the case of the law $dn=0$, we expect this problem to occur generically.

\item\label{rem:noninteract} We have used, as usual in the HBDM, that the wave function $\psi$ is a multi-time wave function, i.e., that it is defined on $\M^N$ or, for a variable number of particles, on $\bigcup_{n=0}^\infty \M^n$, or on the set of spacelike configurations
\be\label{Sdef}
\SC_N=\bigl\{(x_1,\ldots,x_N)\in\M^N: \: \forall j\neq k: x_j \sim x_k\text{ or }x_j=x_k\bigr\}\,,
\ee
where $x\sim y$ means that $x$ is spacelike separated from $y$, or on $\bigcup_{n=0}^\infty \SC_n$. We have also used the divergence-freeness \eqref{divergence}, but it played no role that there was no interaction between the particles. In fact, multi-time wave functions with interaction can be defined on $\bigcup_{n=0}^\infty \SC_n$ \cite{PT13}, with the interaction implemented through the creation and annihilation of particles, and they still satisfy \eqref{divergence} except for additional terms in the probability balance equation that correspond to the creation and annihilation of particles, terms that do not interfere with our analysis. As a consequence, in the case with interaction, the analysis presented in this paper still applies, and the HBDM still works with kinks.
\end{enumerate}

\bigskip

\noindent\textit{Acknowledgments.} 
Both authors acknowledge support from the John Templeton Foundation, grant no.\ 37433.

\end{document}